\documentclass[11pt, a4paper]{article}

\usepackage[utf8]{inputenc}
\usepackage[margin=1in]{geometry}
\usepackage{url}
\usepackage{placeins}
\usepackage{setspace}
\usepackage{amsmath,amssymb,amsfonts}
\usepackage{algorithmic}
\usepackage{graphicx}
\usepackage{textcomp}
\usepackage{amsthm}
\usepackage{mathtools}
\usepackage{algorithm}
\usepackage{stfloats}
\usepackage{verbatim}
\usepackage{array}
\usepackage[colorlinks=true, allcolors=blue]{hyperref} 
\usepackage{cleveref}
\usepackage[caption=false,font=normalsize,labelfont=sf,textfont=sf]{subfig}
\usepackage{balance}

\newcommand{\copyrightnotice}{%
  \thanks{This is a Gold Open Access article made available under the CC-BY license. Published in \textit{Applied Soft Computing}. DOI: \href{https://doi.org/10.1016/j.asoc.2025.114546}{10.1016/j.asoc.2025.114546}}
}

\title{Memetic Covariance Matrix Adaptation Evolution Strategy for Bilinear Matrix Inequality Problems in Control System Design\copyrightnotice}

\author{
    Syue-Cian Lin$^{1}$, 
    Wei-Yu Chiu$^{2,}$\thanks{Corresponding author: weiyu.chiu@deakin.edu.au}, 
    Chien-Feng Wu$^{3}$ \\[1ex]
    \small $^{1}$Department of Electrical Engineering, National Tsing Hua University, Hsinchu 300044, Taiwan \\
    \small $^{2}$School of Information Technology, Faculty of Science, Engineering and Built Environment, \\ \small Deakin University, VIC, Australia \\
    \small $^{3}$Department of Electrical Engineering, National Taipei University, New Taipei City 237, Taiwan
}

\date{} 

\begin{document}

\maketitle

\begin{abstract}
Bilinear Matrix Inequalities (BMIs) are fundamental to control system design but are notoriously difficult to solve due to their nonconvexity. This study addresses BMI-based control optimization problems by adapting and integrating advanced evolutionary strategies. Specifically, a memetic Covariance Matrix Adaptation Evolution Strategy (memetic CMA-ES) is proposed, which incorporates a local refinement phase via a (1+1)-CMA-ES within the global search process. While these algorithmic components are established in evolutionary computing, their tailored integration and specific tuning for control design tasks represent a novel application in this context. Experimental evaluations on $H_{\infty}$ controller synthesis and spectral abscissa optimization demonstrate that the proposed method achieves superior performance compared to existing BMI solvers in terms of both solution quality and robustness. This work bridges the gap between evolutionary computation and control theory, providing a practical and effective approach to tackling challenging BMI-constrained problems.
\end{abstract}

\vspace{1ex}

\section{Introduction}\label{sec_intro}
Control problems are often effectively addressed using linear matrix inequalities (LMIs)~\cite{LMI1,LMI2,LMI3}; however, not all control design tasks can be fully captured within the LMI framework. An alternative and more general formulation is provided by bilinear matrix inequalities (BMIs), which extend the applicability of matrix inequality-based methods. A BMI is a type of constraint in which the matrix depends bilinearly on two sets of variables—meaning that the matrix entries contain products of variables from each set. Unlike LMIs, which are convex and linear in the decision variables, BMIs introduce bilinear terms that increase the complexity of the problem, often resulting in a nonconvex and computationally challenging optimization landscape.

In recent years, BMI-based controller synthesis has attracted increasing research interest~\cite{BMI2,BMI44,BMI6,BMI7,BMI8,BMI9,BMI10}. Compared to LMIs, BMIs offer greater modeling flexibility, enabling the formulation of control problems with structured, decentralized, or nonlinear constraints. This makes BMI formulations particularly suitable for tackling complex control scenarios involving uncertainty and nonlinearity~\cite{00vanantwerp}. Moreover, BMI-based methods can often yield less conservative controller designs than their LMI counterparts~\cite{BMI_less_conservative}, thus providing improved performance and robustness in practical applications.

While various BMI solution methods have been extensively studied, many of the existing approaches exhibit notable limitations. Some techniques require problem-specific preprocessing. For instance, the convex-concave decomposition method (CCDM)~\cite{1206dinh} requires preliminary decomposition and linearization steps, while the inner-convex approximation method (ICAM)~\cite{1212dinh} relies on problem approximations before application. Other solution techniques, such as alternating minimization and path-following methods~\cite{0809ostertag,0807ostertag,783595}, are often tailored to specific problem types and may necessitate reformulating the BMI structure when applied to different scenarios. Additionally, optimization tools like HIFOO~\cite{11arzelier,18conti,hifoo}, LMIRank~\cite{orsilmirank}, and PENBMI~\cite{05henrion,penbmiuserguide} are typically prone to convergence to local optima, which limits their effectiveness for highly nonconvex BMI formulations.

To overcome some of these limitations, the method of reduction of variables (MRVs)~\cite{17chiu} was proposed to transform BMI problems into LMI problems without relying on specific structural assumptions. MRVs employ a hybrid multiobjective immune algorithm (HMOIA) with stochastic search mechanisms to explore the global solution space. However, MRVs require predefined upper and lower bounds for controller gain matrix entries or closed-loop eigenvalues. These static bounds may not generalize well across diverse BMI problem settings, potentially restricting the method’s ability to locate the true global optimum.

An alternative approach, the quality diversity optimization method (QDOM)~\cite{QDOM}, avoids explicitly bounding controller gains by incorporating a bound-searching algorithm that dynamically adjusts the search range. While this enables global exploration, it introduces additional complexity and computational cost. Moreover, QDOM may produce solutions with excessively large feedback gains—an undesirable feature in controller synthesis that can compromise stability or practicality. These challenges underscore the need for more refined strategies that can both regulate feedback magnitudes and maintain feasible, stable control performance.

\begin{table}[htbp]
\label{tab:method-comparison}
\centering
\caption{Comparison of Representative BMI Solution Methods}
\resizebox{\textwidth}{!}{%
\begin{tabular}{|p{4.8cm}|p{5.6cm}|p{5.6cm}|}
\hline
\textbf{Approach} & \textbf{Pros} & \textbf{Cons} \\
\hline
Convex--Concave Decomposition Method (CCDM)~\cite{1206dinh} &
Handles certain non‑convex structures; relatively fast for structured problems &
Requires decomposition and linearization; sensitive to problem structure \\
\hline
Inner Convex Approximation Method (ICAM)~\cite{1212dinh} &
Iteratively approximates non‑convex problems &
Needs problem reformulation; guarantees only local convergence \\
\hline
HIFOO~\cite{11arzelier,18conti,hifoo} &
Ready‑to‑use MATLAB toolbox for $H_\infty$ optimization &
Prone to local optima; may struggle with large-scale or stiff problems \\
\hline
LMIRank~\cite{orsilmirank} &
Applies rank minimization heuristics; integrates with LMI tools &
Problem-specific formulation needed; local solutions only \\
\hline
PENBMI~\cite{05henrion,penbmiuserguide} &
General-purpose BMI solver; supports nonlinear constraints &
Can be slow; limited success in hard BMI benchmarks \\
\hline
Method of Reduction of Variables (MRVs)~\cite{17chiu} &
Transforms BMI to LMI; uses evolutionary global search &
Requires predefined variable bounds; less adaptable across problems \\
\hline
Quality Diversity Optimization Method (QDOM)~\cite{QDOM} &
Global search without static bounds; good solution diversity &
High computational cost; may yield excessively large feedback gains \\
\hline
\textbf{Proposed memetic CMA‑ES} &
Adaptive search without bounds; balances global and local exploration; regulates gain size &
Higher algorithmic complexity than single‑phase approaches \\
\hline
\end{tabular}}
\end{table}

Unlike QDOM, which was specifically developed for BMI-based control system design~\cite{QDOM}, the covariance matrix adaptation evolution strategy (CMA-ES)~\cite{CMAES} is a general-purpose optimization algorithm.
Despite the absence of a universal theoretical guarantee for global convergence, CMA-ES has exhibited strong empirical performance across a wide range of continuous, nonconvex optimization problems~\cite{hansen2016cma}.
CMA-ES features a self-adaptive mechanism for navigating complex landscapes, adjusting its search distribution via an evolving covariance matrix and adaptive step-size control. The evolution path, constructed from a history of successful steps, informs both the scaling and orientation of the sampling distribution, enabling efficient exploration without requiring predefined variable bounds. While CMA-ES has not been widely applied to BMI-constrained control synthesis, its inherent adaptability makes it a promising candidate for this domain when appropriately modified and integrated with problem-specific mechanisms.

To improve exploitation while preserving robust exploration, this work introduces a memetic optimization framework—termed memetic CMA-ES—that integrates global and local search mechanisms in a problem-specific manner to address the inherent nonconvexity and numerical challenges of BMI-constrained control problems. The first phase employs the CMA-ES, which adaptively adjusts its sampling distribution and step size to explore high-dimensional, rugged search spaces without requiring predefined variable bounds. To enhance solution quality and feasibility recovery, a second phase based on the (1+1)-CMA-ES~\cite{(1+1)-CMAES} is introduced for localized refinement. This variant emphasizes adaptive step-size control and focused exploitation, making it well-suited for fine-tuning candidate solutions in constrained control settings.

While CMA-ES has previously been combined with local refinement heuristics in general optimization contexts, such combinations often lack the domain-specific adaptations required for BMI-constrained controller design. In contrast, the proposed two-phase memetic structure explicitly addresses BMI-specific challenges. It integrates a penalty-guided constraint-handling mechanism that dynamically discourages excessive feedback gains—an issue commonly overlooked in previous works. This structured coordination between exploration and exploitation offers improved convergence reliability, feasibility, and control relevance. Table~\ref{tab:method-comparison}  compares the aforementioned representative BMI solution methods.

To evaluate the effectiveness of the proposed approach, a comprehensive benchmark test was conducted on BMI-constrained control problems from the COMPleib library~\cite{04leibfritz11}.
In terms of achieving the best $H_\infty$-norm values, the success rates of various methods were as follows: HIFOO (10.64\%), PENBMI (0\%), CCDM (10.64\%), MRVs (21.28\%), QDOM (17.02\%), and the proposed memetic CMA-ES (85.11\%).
For the problem of spectral abscissa optimization, which concerns minimizing the largest real part of the closed-loop system eigenvalues, the observed success rates were: HIFOO (16.67\%), LMIRank (20\%), PENBMI (26.67\%), CCDM (16.67\%), ICAM (20\%), MRVs (23.33\%), QDOM (40\%), and the proposed memetic CMA-ES achieving the highest rate at 73.33\%.
These results demonstrate that memetic CMA-ES significantly outperforms existing approaches, both in terms of solution quality and consistency across diverse BMI problems.

This article contributes to the literature by introducing a problem-specific memetic optimization framework that strategically combines global and local search mechanisms to address the unique challenges of BMI-constrained controller design. In contrast to previous works that have paired CMA-ES with generic local heuristics, our method offers three key innovations: (i) a structured two-phase design that separates global exploration and feasibility-driven local refinement, (ii) a (1+1)-CMA-ES refinement stage specifically adapted to recover feasibility in BMI formulations, and (iii) a penalty-guided mechanism that regulates feedback gain magnitudes without relying on predefined variable bounds. These algorithmic enhancements are tightly aligned with the structure and constraints of BMI problems, ensuring both solution quality and practical relevance. Rather than proposing a new standalone algorithm, this work delivers a reproducible and scalable strategy that outperforms state-of-the-art solvers across diverse benchmark problems. The significant performance gains demonstrated in our experiments validate the efficacy of this domain-aware memetic approach.

The remainder of this paper is organized as follows. Section~\ref{sec:relatedworks} reviews recent developments related to nonlinear systems, fuzzy control, and evolutionary optimization that provide a broader context for this study. Section~3 introduces the problem formulation. The proposed memetic CMA-ES algorithm is detailed in Section~4. Section~5 presents the experimental results and comparative performance analysis. Finally, Section~6 concludes the paper and outlines possible directions for future work.

\section{Related Works}
\label{sec:relatedworks}

In addition to methods directly targeting BMI-constrained controller design, recent progress in related areas such as fuzzy control, nonlinear systems with time delay, and advanced optimization algorithms offers valuable insights. For instance, Shamrooz Aslam et al.~\cite{SHAMROOZASLAM2023119204} proposed a Takagi–Sugeno fuzzy control method for time-delay systems using an extended sliding mode observer, which demonstrates robustness under measurement uncertainty. Other works by the same author~\cite{aslam2022stability,aslam2021design} contribute to the analysis of admissibility in descriptive systems and sliding mode control under actuator/sensor faults.

From an optimization standpoint, Aljaidi et al.~\cite{aljaidi2025two} introduced a two-phase differential evolution algorithm with perturbation and covariance matrix adjustments for parameter estimation in proton exchange membrane fuel cells. While not directly aimed at BMI problems, their integration of adaptive strategies and metaheuristic design principles aligns with the motivation behind evolutionary frameworks like the proposed memetic CMA-ES.

These studies underscore the value of integrating control theory with modern optimization techniques, particularly in handling uncertainty, system complexity, and nonlinearities.

\section{Problem Formulation}\label{sec_model}

The design of the system and controller is formulated as a BMI optimization problem in this section, commonly approached via $H_{\infty}$ optimization. 
The system under research~\cite{QDOM,17chiu} has the following form:
\begin{equation}\label{eq_sys_dynamics}
\left\{
\begin{array}{l}
\boldsymbol{\dot{x}=Ax+B_1w+Bu} \\
\boldsymbol{z=C_1x+D_{11}w+D_{12}u} \\
\boldsymbol{y=Cx} \\
\end{array}
\right.
\end{equation}
where $\boldsymbol{x} \in \mathbb{R}^{n_x}$ denotes the state vector, $\boldsymbol{w} \in \mathbb{R}^{n_w}$ is the vector of exogenous inputs (e.g., disturbances and noise), $\boldsymbol{u} \in \mathbb{R}^{n_u}$ is the control input, $\boldsymbol{y} \in \mathbb{R}^{n_y}$ is the measured output available for feedback, and $\boldsymbol{z} \in \mathbb{R}^{n_z}$ is the performance output to be regulated.

The control objective is achieved using a static output feedback controller, where the control input $\boldsymbol{u}$ is a linear function of the measured output $\boldsymbol{y}$:
\begin{equation}
\label{eq:controller}
\boldsymbol{u} = \boldsymbol{Fy} = \boldsymbol{FCx}
\end{equation}
where $\boldsymbol{F} \in \mathbb{R}^{n_u \times n_y}$ is the constant controller gain matrix to be designed.

Substituting the control law \eqref{eq:controller} into the system dynamics \eqref{eq_sys_dynamics} yields the closed-loop system from the exogenous input $\boldsymbol{w}$ to the performance output $\boldsymbol{z}$:
\begin{equation}
\label{eq:closed_loop_system}
\left\{
\begin{aligned}
\dot{\boldsymbol{x}} &= (\boldsymbol{A+BFC})\boldsymbol{x} + \boldsymbol{B_1w} \equiv \boldsymbol{A_F x} + \boldsymbol{B_1w} \\
\boldsymbol{z} &= (\boldsymbol{C_1+D_{12}FC})\boldsymbol{x} + \boldsymbol{D_{11}w} \equiv \boldsymbol{C_F x} + \boldsymbol{D_{11}w}
\end{aligned}
\right.
\end{equation}
Here, $\boldsymbol{A_F}$ and $\boldsymbol{C_F}$ are the closed-loop system matrices, which are dependent on the control gain $\boldsymbol{F}$.

The dynamics of this closed-loop system can be described by the transfer function matrix $\boldsymbol{G}_{\text{cl}}(s)$ from $\boldsymbol{w}$ to $\boldsymbol{z}$. This transfer function has the following state-space realization:
\begin{equation}
\boldsymbol{G}_\text{cl}(\boldsymbol{F}) =
\left[
\begin{array}{c|c}
\boldsymbol{A_F} & \boldsymbol{B_{1}} \\
\hline
\boldsymbol{C_F} & \boldsymbol{D_{11}}
\end{array}
\right]
= \boldsymbol{C_F}(s\boldsymbol{I} - \boldsymbol{A_F})^{-1} \boldsymbol{B_1} + \boldsymbol{D_{11}}.
\end{equation}

The static output feedback $H_\infty$ control problem is to find a controller gain $\boldsymbol{F}$ that stabilizes the closed-loop system and minimizes the $H_\infty$ norm of this transfer function. The $H_\infty$ norm is defined as the peak gain of the system's frequency response:
\begin{equation}
    \| \boldsymbol{G}_{\text{cl}} \|_{\infty} = \sup_{\omega \in \mathbb{R}} \sigma_{\max} (\boldsymbol{G}_{\text{cl}}(j\omega)),
\end{equation}
where $\sigma_{\max}(\cdot)$ denotes the maximum singular value of a matrix.

Physically, the $H_\infty$ norm represents the worst-case energy gain from the exogenous inputs $\boldsymbol{w}$ to the performance outputs $\boldsymbol{z}$. Minimizing this norm enhances system robustness by guaranteeing the attenuation of disturbances and ensuring stability and performance in the presence of uncertainties.

The $H_\infty$ optimization problem can be formulated as follows~\cite{03hoi,QDOM}:
\begin{equation}
\label{Hinf definition}
	\begin{array}{rl}
    		\displaystyle\min_{\boldsymbol{F},\gamma,\boldsymbol{P}} & ||\boldsymbol{G}_\text{cl}||_{\infty} \\
    		\mbox{subject to} &
    		\left[
		\begin{array}{ccc}
		\boldsymbol{PA_F} \boldsymbol{+A_F^{\boldsymbol{^\top}}} \boldsymbol{P} & \boldsymbol{PB_1} & \boldsymbol{C}_\textbf{F}^{\boldsymbol{\top}}\\
		(\boldsymbol{PB_1})^T & -\gamma \boldsymbol{I} & \boldsymbol{D_{11}^\top} \\
		\boldsymbol{C_F} & \boldsymbol{D_{11}} & -\gamma \boldsymbol{I}
		\end{array}
		\right]
		< 0 \\
		&
		\gamma > 0, \boldsymbol{P} > 0.
    \end{array}
\end{equation}

The control design subject to the BMI constraint in (\ref{Hinf definition}) can be compactly expressed as
\begin{equation}
\label{BMI problems}
	\begin{array}{rl}
		\displaystyle\min_{\boldsymbol{\alpha},\boldsymbol{X}} & \mathcal{F}(\boldsymbol{\alpha}) \\
    		\mbox{subject to} & \mathcal{BMI}(\boldsymbol{\alpha},\boldsymbol{X}) < 0
	\end{array}
\end{equation}
where 
 $\mathcal{F}(\boldsymbol{\alpha})=||\boldsymbol{G}_\text{cl}||_{\infty} $
and
\begin{equation}
\label{BMI123456}
\mathcal{BMI}(\boldsymbol{\alpha},\boldsymbol{X}) < 0
\end{equation}
denotes the BMI constraint; 
$\boldsymbol{\alpha}$ represents all the
external variables  and $\boldsymbol{X}$ represents all the internal variables. In (\ref{Hinf definition}), the external variable is $\boldsymbol{\alpha}=\boldsymbol{F}$, while the remaining variables—$\boldsymbol{P}$ and $\gamma$—are internal variables $\boldsymbol{X}=(\gamma,\boldsymbol{P})$. 
Once a feasible solution to the BMI constraint is obtained, it implies that the closed-loop system matrix \(\boldsymbol{ A + BFC} \) admits a Lyapunov function \( V(\boldsymbol{x}) = \boldsymbol{x^T P x} \) satisfying \( \boldsymbol{A_F^T P + P A_F} < 0 \). This condition guarantees exponential stability of the closed-loop system, provided that the disturbance \(\boldsymbol{w}\) is absent or sufficiently bounded.

\begin{figure}[ht]
    \centering
    \includegraphics[width=0.6\linewidth]{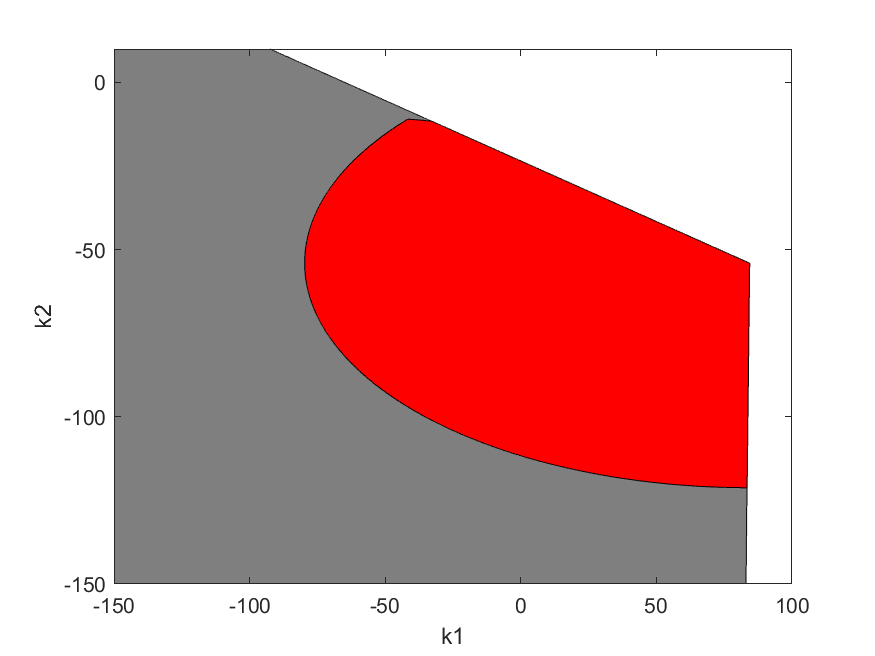}
    \caption{Illustration of the feasible region and objective value distribution for the BMI benchmark AC4, where $k_1$ and $k_2$ denote the two entries of the controller gain. The red region indicates the set of points for which the objective value in (\ref{BMI problems2_ori}) is less than 1, while the overall feasible region comprises both the red and gray areas.
}
    \label{fig:AC4fig2}
\end{figure}

Unfortunately, directly solving~(\ref{BMI problems}) can result in excessively large controller gains.
Fig.~\ref{fig:AC4fig2}
shows the feasible region
of the BMI benchmark problem AC4; the controller gains $k_1$ and $k_2$ become unbounded in certain directions. This issue arises because the \(H_\infty\) optimization in~(\ref{Hinf definition}) does not inherently impose constraints on the magnitude of controller gains within the BMI formulation. Consequently, it may lead to impractically large feedback gains. To ensure practical applicability, such excessive gains must be avoided.

To solve this issue, a penalty term on controller gains is introduced, restricting their growth and preventing instability or unrealistic control behavior, as commonly applied in solution solvers~\cite{05henrion,BURKE2006339,penbmiuserguide}. The BMI problem can then be reformulated as follows: 
\begin{equation}
\label{BMI problems2_ori}
	\begin{array}{rl}
		\displaystyle\min_{\boldsymbol{\alpha},\boldsymbol{X}} & \mathcal{F}(\boldsymbol{\alpha}) + \beta \|\boldsymbol{\alpha}\|_2 \\
    		\text{subject to} & \mathcal{BMI}(\boldsymbol{\alpha},\boldsymbol{X}) < 0. \\
	\end{array}
\end{equation}
where $\beta>0$   balances the objective value and the 2-norm of the feedback gain, i.e., $\|\boldsymbol{\alpha}\|_2$.  The penalty restricts the algorithm from exploring excessively large feedback gains.

While this study primarily focuses on the $H_{\infty}$ control framework, it is important to note that many other critical control objectives can be formulated as BMIs. A prominent example is the optimization of the spectral abscissa for system stabilization. Given that spectral abscissa problems share the same underlying nonconvex structure as the $H_{\infty}$ synthesis described above, they serve as an additional benchmark for BMI solution methods. 
A detailed treatment of spectral abscissa optimization in the BMI framework can be found in~\cite{17chiu}.

\section{Proposed Memetic CMA-ES}\label{sec_method}

To solve~(\ref{BMI problems2_ori}), 
Algorithm~\ref{algorithm1} presents the memetic CMA-ES framework, which integrates a global search using standard CMA-ES and a deterministic local refinement via (1+1)-CMA-ES presented in Algorithm~\ref{algorithm2}. To elaborate on the proposed memetic CMA-ES in Algorithm 1, we sequentially discuss its major components: initialization, sampling, local search integration via Algorithm 2, and adaptive updates. We then proceed to detail Algorithm 2. Finally, the computational complexity is analyzed using Big-O notation.

The variable reduction method in~\cite{17chiu} can be further used to transform~(\ref{BMI problems2_ori}) into 
\begin{equation}
\label{problem with one decision variable}
	\begin{array}{rl}
		\displaystyle\min_{\boldsymbol{\alpha}} &  \mathcal{F}(\boldsymbol{\alpha}) + \beta \|\boldsymbol{\alpha}\|_2 \\
    		\mbox{subject to} & \mathcal{\lambda}^*(\boldsymbol{\alpha}) < 0. \\
	\end{array}
\end{equation}
where $\mathcal{\lambda}^*(\boldsymbol{\alpha})$ represents an eigenvalue problem (a convex problem) that can solved using interior-point methods. 
It can be shown that 
if  $\boldsymbol{\alpha}$ with $\mathcal{\lambda}^*(\boldsymbol{\alpha}) < 0$ exists, then~(\ref{BMI problems2_ori}) is feasible. This property will be used to check the feasibility  of points generated during the algorithm iterations; if a point is infeasible,
 a large penalty is added to deprioritize it during selection.

We further define the fitness function as 
\begin{equation}
\label{fitness_function}
    f_{\text{BMI}}(\boldsymbol{\alpha}) = -\mathcal{F}(\boldsymbol{\alpha}) - \beta \|\boldsymbol{\alpha}\|_2.
\end{equation}
It is worth noting that negative signs were introduced to convert the original minimization objective into a maximization form, aligning with the common convention that fitness functions are typically defined for maximization.

\textbf{Initialization.}
The optimization process in Algorithm~\ref{algorithm1} begins by initializing the external variable \(\boldsymbol{\alpha}\), which corresponds to the controller gain matrix \(\boldsymbol{F}\) as defined in the previous section. To maintain consistency, \(\boldsymbol{\alpha}\) is structured as a column vector throughout the algorithm. In this setting,  \(t_{\text{max}}\) represents the maximum number of generations within each iteration.

\begin{algorithm}[H]
\caption{Memetic CMA-ES}
\label{algorithm1}
\begin{algorithmic} [1]
    \REQUIRE $t_{\text{max}},\boldsymbol{m}^{(t)}$ \\
    \hspace{-1.7em}\textbf{Initialize:} $t, p,\mu,\sigma^{(t)} = 0.3, \boldsymbol{\Sigma}^{(t)} = \boldsymbol{I}, \boldsymbol{p}^{(t)}_{\sigma} = \boldsymbol{0}, \boldsymbol{p}^{(t)}_{\text{cov}} = \boldsymbol{0}$ 
    \ENSURE $f^{\text{best}}_{\text{BMI}}(\boldsymbol{\alpha}^{(t)}_i)$
    \\
        \WHILE{$t \neq t_{\text{max}}$}
                \FOR{$i = 1$ \TO $p$}
                    \STATE Generate $\boldsymbol{\alpha}^{(t)}_i$ with multivariate normal distribution and calculate the fitness $f_{\text{BMI}}(\boldsymbol{\alpha}^{(t)}_i)$
                    \STATE Refine the local region in $\boldsymbol{\alpha}^{(t)}_i$ by $f_{\text{BMI}}(\boldsymbol{\alpha}^{(t)}_i)$ using Algorithm 2
                    \STATE $t = t + 1$
                \ENDFOR
                \STATE \text{Sort by the fitness value of each generation}
                \STATE Store the best solution into $f^{\text{best}}_{\text{BMI}}(\boldsymbol{\alpha}^{(t)}_i)$
                \STATE $ \text{Update the mean }$

                \STATE $ \text{Update the evolution paths}$
                \STATE $ \text{Update the covariance matrix}$
                \STATE $ \text{Update the step size}$
                \STATE $ \text{Perform parameter reset and eigenvalue correction}$
        \ENDWHILE
\end{algorithmic}
\end{algorithm}

\textbf{Sampling.}
In each generation of Algorithm~\ref{algorithm1}, the global search component—standard CMA-ES—samples \(p\) candidate solutions \(\boldsymbol{\alpha}^{(t)}_i\) from a multivariate normal distribution centered at the mean \(\boldsymbol{m}^{(t)}\), with step size \(\sigma^{(t)}\) and covariance matrix \(\boldsymbol{\Sigma}^{(t)}\):

\begin{equation}
\label{normal distribution}
    \boldsymbol{\alpha}^{(t)}_i = \boldsymbol{m}^{(t)} + \sigma^{(t)} \boldsymbol{\xi}^{(t)}.
\end{equation}
where \(\boldsymbol{\xi}^{(t)} \sim\mathcal{N}(\boldsymbol{0},\boldsymbol{\Sigma}^{(t)}\)), 
Each candidate \(\boldsymbol{\alpha}^{(t)}_i\) is then evaluated using the BMI-specific fitness function  $f_{\text{BMI}}(\boldsymbol{\alpha}^{(t)}_i )$.
 If  \(\boldsymbol{\alpha}^{(t)}_i \) is infeasible, a large  penalty (e.g., \(10^5\)) is added to its fitness to deprioritize it during selection.

\textbf{Local Search Integration.} After generating each offspring \(\boldsymbol{\alpha}^{(t)}_i\) via standard CMA-ES in line~3 of Algorithm~\ref{algorithm1}, a local search is immediately invoked using Algorithm~\ref{algorithm2}. Specifically, line~4 of Algorithm~\ref{algorithm1} passes \(\boldsymbol{\alpha}^{(t)}_i\) as the input to Algorithm~\ref{algorithm2}, which performs (1+1)-CMA-ES-based refinement. The locally optimized solution \(\boldsymbol{\alpha}_{\text{opt}}\) is then used to replace the original offspring prior to the population sorting in line~6. This deterministic and consistent application of local search after each global step ensures improved exploitation of the local region and enhances convergence quality.

\begin{figure}[p]
    \centering
    \begin{equation*}
    \begin{array}{cc}
        \includegraphics[width=0.5\textwidth]{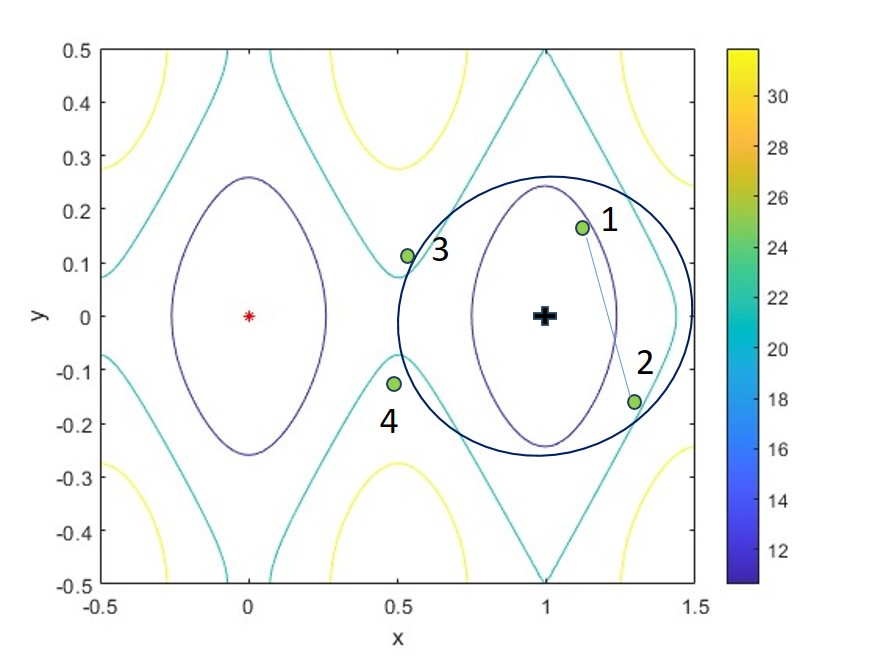} & \includegraphics[width = 0.5\textwidth]{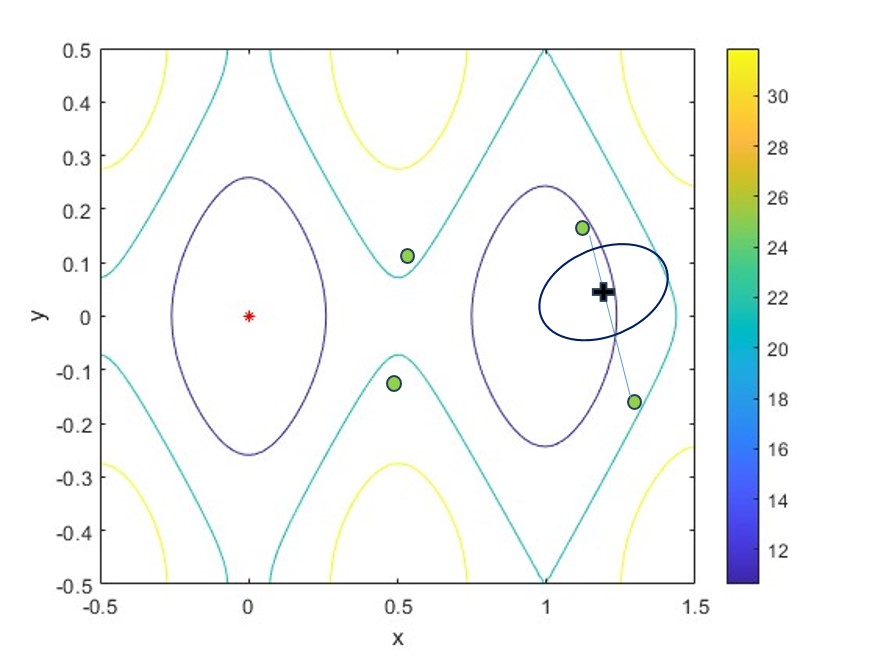}\\
        \mbox{(a)} & \mbox{(b)}\\
        \includegraphics[width=0.5\textwidth]{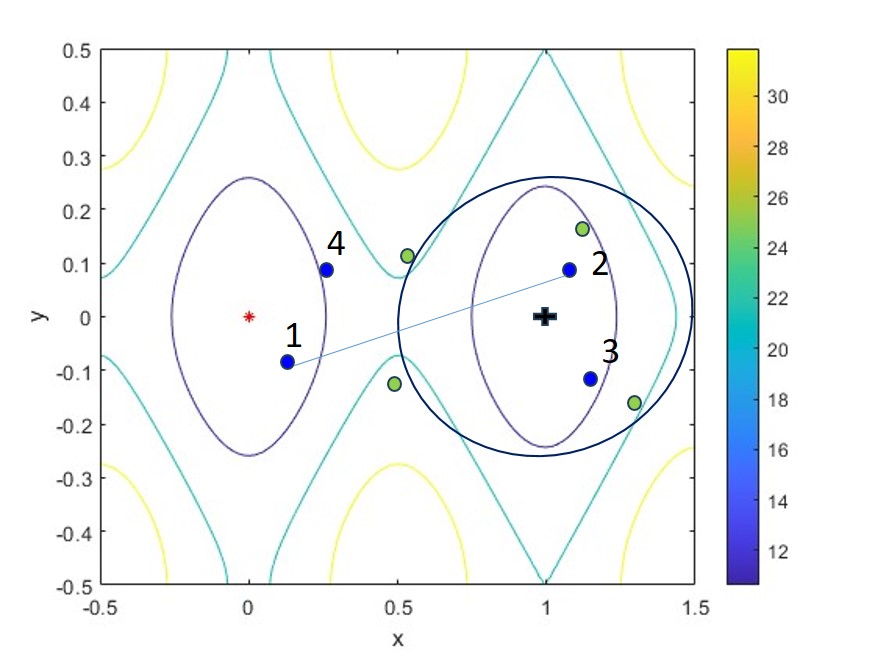} & \includegraphics[width = 0.5\textwidth]{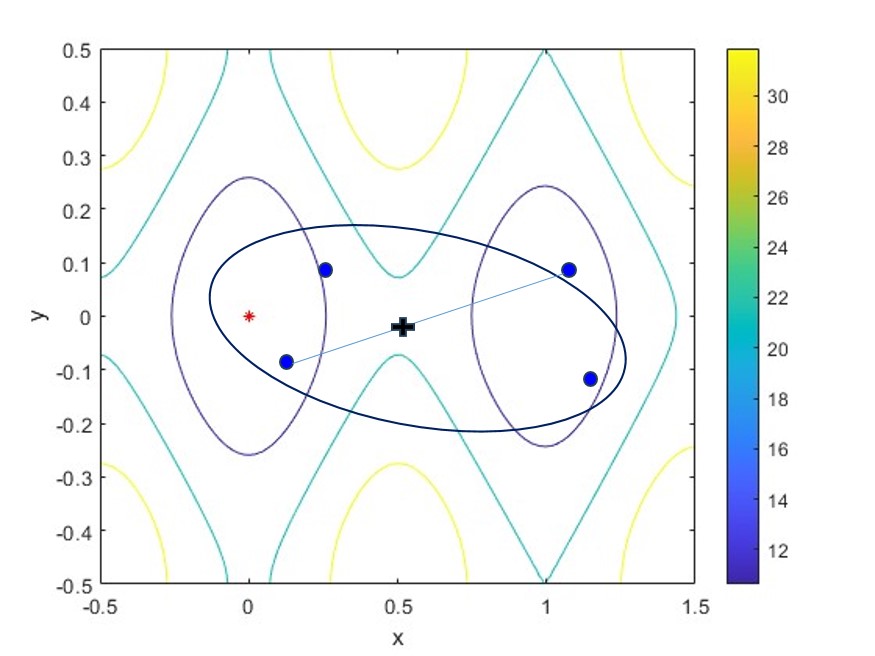}\\
        \mbox{(c)} & \mbox{(d)}\\
    \end{array}
    \end{equation*}
    \caption{Visualization of population dynamics in CMA-ES and memetic CMA-ES. The black crosses are the mean in the algorithms, the black circles are the one standard deviation of the covariance matrix, the green points are the offsprings in current generation, the blue points are the better solution than original offsprings (green points) from memetic CMA-ES, the red cross-star is the global optimal point, and the contour value is from the right side color bar. There is a number besides the blue points. The smaller the number, the greater the objective value. The blue line is the baseline of mean, and the required mean for the next generation will be on the blue line, influenced by weighted contributions. (a) The standard CMA-ES generates the children with the mean and covariance matrix. (b) The next generation of the mean and covariance matrix in standard CMA-ES. (c) The memetic CMA-ES generates the children with the mean and covariance matrix. (d) The next generation of mean and covariance matrix in memetic CMA-ES. This shows how the refinement of (1+1)-CMA-ES allows the memetic CMA-ES to lead the global search in a more appropriate direction.}
    \label{fig:exfig3}
\end{figure}

The (1+1)-CMA-ES serves as a post-processing operator that explores the neighborhood of each offspring using a smaller step size—specifically, one-tenth of the global \(\sigma^{(t)}\)—and an identity covariance matrix to enforce unbiased and fine-grained local perturbations. This mechanism complements the global search by correcting possible missteps caused by the covariance adaptation in standard CMA-ES, as illustrated in Fig.~\ref{fig:exfig3}(a)(b). The integration of both search components allows a balanced trade-off between exploration and exploitation, promoting stable and reproducible convergence.

It is worth mentioning that, in addition to the (1+1)-CMA-ES used in our framework, other local search methods such as gradient-based approaches and the Nelder--Mead algorithm are also commonly considered in optimization. However, in this study, we do not assume the differentiability of the objective function. For instance, $H_\infty$ optimization involves taking the supremum over frequency, which introduces non-smooth behavior. Moreover, the maximum singular value function is inherently non-differentiable when two singular values are close, leading to abrupt changes in gradients. Even when using automatic differentiation libraries, differentiating through eigenvalue computations or frequency sweeps is unreliable. As a result, gradient-based local search methods are not appropriate in this context.

Regarding Nelder--Mead, although it is a derivative-free method, it operates by iteratively transforming a simplex, which can be effective in low-dimensional, smooth problems. However, in higher-dimensional settings or for objective functions with flat regions and sharp ridges---as is often the case in controller tuning and $H_\infty$ problems---it becomes inefficient and may fail to converge. In contrast, (1+1)-CMA-ES adaptively shapes its search distribution through the covariance matrix, allowing it to navigate complex landscapes more effectively by following valleys, ridges, and non-separable structures in the objective function.

\textbf{Mean Update.} In line~7 of Algorithm~\ref{algorithm1}, the objective values across generations are sorted in descending order to identify the $\mu$ best offspring for the following computation. 
To simplify the notation, we use the same symbol $\boldsymbol{\alpha}_{i}^{(t)}$ to represent individuals both before and after sorting by fitness values.
Meanwhile, if the fitness value $f_{\text{BMI}}(\boldsymbol{\alpha}^{(t)}_i)$ is greater than $f^{\text{best}}_{\text{BMI}}(\boldsymbol{\alpha}^{(t)}_i)$, it is stored as $f^{\text{best}}_{\text{BMI}}(\boldsymbol{\alpha}^{(t)}_i)$ to retain the best result in line~8 of Algorithm~1. Based on this order, the mean 
\(\boldsymbol{m}^{(t+1)}\)
will be adjusted through evolution in line~9 of Algorithm~\ref{algorithm1}:
\begin{equation}
    \boldsymbol{m}^{(t+1)} =\sum^{\mu}_{i = 1} \theta_i \boldsymbol{\alpha}^{(t)}_i
\end{equation}
where the $\theta_i$ is the weighted average of $\mu$ selected from the samples $\boldsymbol{\alpha}^{(t)}_1 ,\dots, \boldsymbol{\alpha}^{(t)}_p$, and generally $\mu$ is the half of $p$. 
For a detailed value of $\theta_{i}$, please refer to the Appendix.
 
\textbf{Evolution Paths Update.}
In the process of optimizing the objective value, the rank-one-update and rank-$\mu$-update mechanisms enhance the adaptability and precision of the search. Specifically, the rank-$\mu$-update method selects the top $\mu$ individuals from the current population to update the covariance matrix, allowing the algorithm to adjust more efficiently toward the optimal search direction. Meanwhile, the rank-one-update incrementally modifies the evolution path based on a single selected step in each generation, enabling the algorithm to accumulate directional information over multiple generations. This contributes to a more stable and consistent long-term search.

The parameter $\boldsymbol{p}^{(t)}_{\text{cov}}$ represents the evolution path for the covariance matrix, while $\boldsymbol{p}^{(t)}_{\sigma}$ corresponds to the evolution path for the step size. Both $\boldsymbol{p}^{(t)}_{\text{cov}}$ and $\boldsymbol{p}^{(t)}_{\sigma}$ are vectors. In the rank-one-update step, as implemented in line~10 of Algorithm~\ref{algorithm1}, the evolution paths are defined as follows:

\begin{align}
    \boldsymbol{p}^{(t+1)}_{\sigma} &= (1 - c_{\sigma})\boldsymbol{p}^{(t)}_{\sigma} + 
    \sqrt{c_{\sigma}(2-c_{\sigma})\mu_{\text{eff}}} \, ((\mathbf{\Sigma}^{(t)})^{-\frac{1}{2}}) \, \frac{\boldsymbol{m}^{(t+1)} - \boldsymbol{m}^{(t)}}{\sigma^{(t)}}
    \label{function19}\\
    \boldsymbol{p}^{(t + 1)}_{\text{cov}} &= (1 - c_{\text{c}})\boldsymbol{p}^{(t)}_{\text{cov}} +
    h_{\sigma}\sqrt{c_c(2-c_{\text{c}})\mu_{\text{eff}}} \frac{\boldsymbol{m}^{(t+1)} - \boldsymbol{m}^{(t)}}{\sigma^{(t)}} \label{function18}
\end{align}
where $c_{\sigma}$ is the decay rate for the step size, and $\mu_{\text{eff}}$ is the effective sample size of the selected samples, the term $(\mathbf{\Sigma}^{(t)})^{-\frac{1}{2}}$ refers to the square root of the covariance matrix in~\cite{hansen2023cmaevolutionstrategytutorial}, $c_{\text{c}}$ is the decay rate associated with the rank-one update of the covariance matrix, and the Heaviside function \( h_\sigma \) is used to control $ p^{(t+1)}_\text{cov} $ and $ \boldsymbol{\Sigma}^{(t+1)}$, and it is updated before the update of $\boldsymbol{p}^{(t + 1)}_{\text{cov}}$. For a detailed explanation of $(\mathbf{\Sigma}^{(t)})^{-\frac{1}{2}}$, $c_c$, $c_\sigma$, $h_\sigma$ and $\mu_{\text{eff}}$, please refer to the Appendix.

\textbf{Covariance Matrix and Step Size Updates.}
The covariance matrix \(\boldsymbol{\Sigma} \) and step-size \(\sigma\) in lines~11 and 12 of Algorithm~\ref{algorithm1} can be updated by

\begin{align}
\boldsymbol{\Sigma}^{(t+1)} &= 
\begin{aligned}[t]
    &(1 - c_1 - c_\mu \sum_{i=1}^\mu \theta_i) \boldsymbol{\Sigma}^{(t)}
    + c_1 (\boldsymbol{p}_{\text{cov}}^{(t+1)} {\boldsymbol{p}_{\text{cov}}^{(t+1)}}^{\top}
    + \\&(1 - h_{\sigma})c_c(2-c_c)\boldsymbol{\Sigma}^{(t)})
    + \underbrace{c_\mu \sum_{i=1}^\mu \theta_i \frac{\left(\boldsymbol{\alpha}_{i}^{(t)} - \boldsymbol{m}^{(t)}\right)}{\sigma^{(t)}} 
    \frac{\left(\boldsymbol{\alpha}_{i}^{(t)} - \boldsymbol{m}^{(t)}\right)^{\top}}{\sigma^{(t)}}}_{\text{rank-$\mu$-update}}
\end{aligned} \label{covariance} \\
\sigma^{(t+1)} &= \sigma^{(t)} \exp \left( \frac{c_\sigma}{d_\sigma} \left( \frac{\|\boldsymbol{p}_\sigma^{(t+1)}\|_2}{\mathbb{E}\|\mathcal{N}(\mathbf{0}, \mathbf{I})\|_2} - 1 \right) \right) .\label{step size}
\end{align}
The parameters $c_1$ and $c_{\mu}$ represent the learning rates for the rank-one and rank-$\mu$ updates, respectively, and they determine the covariance matrix adaptation in~(\ref{covariance}).
The step size update is controlled by the damping parameter $d_{\sigma}$, and $\mathcal{N}(\mathbf{0}, \mathbf{I})$ denotes the standard multivariate normal distribution. In memetic CMA-ES, the step size governs isotropic exploration, balancing global search in early iterations with refined search later. Meanwhile, the covariance matrix adapts to parameter correlations, guiding the search towards promising regions of $\boldsymbol{\alpha}$. For detailed values of $c_1$, $c_\mu$, and $d_\sigma$, refer to the Appendix.

 To prevent divergence in step size updates in line~13, we reset key parameters—including the covariance matrix, evolution path, mean, and step size—to their initial values. To maintain a positive definite covariance matrix, we apply an eigenvalue correction~\cite{HIGHAM1988103}. This ensures stability without the overhead of Frobenius norm optimization while keeping the matrix well-conditioned for computations in~(\ref{function19}).

\textbf{Details of Algorithm~\ref{algorithm2}.} The local search procedure is elaborated in Algorithm~\ref{algorithm2}, where \(\boldsymbol{\alpha}_{\text{parent}}\), initialized as \(\boldsymbol{\alpha}^{(t)}_i\), represents the parent solution. Lines 2–4 describe the sampling of a local offspring using the identity covariance matrix and the reduced step size. A success indicator \(v_{\text{succ}}\) is set to 1 if the new offspring yields a better fitness value, and 0 otherwise. This affects the update of the success rate \(p_{\text{s}}\) in line 5.

\begin{algorithm}[H]
\caption{(1+1)-CMA-ES}
\label{algorithm2}
\begin{algorithmic} [1]
    \REQUIRE $t_s,\boldsymbol{\alpha}_{\text{parent}},f_{\text{BMI}}(\boldsymbol{\alpha}_{\text{parent}})$\\
    \hspace{-1.7em}\textbf{Initialize:} $v_\text{succ},\sigma_\text{loc}, p_\text{s},p_\text{t}, p^t_{\text{s}}, c_{\text{p}}, c_{\text{cov}}, \boldsymbol{\Sigma} = \boldsymbol{I}, \boldsymbol{p}_\text{c} = \mathbf{0}$, $f^{\text{best}}_{\text{BMI}}(\boldsymbol{\alpha}_{\text{opt}})=f_{\text{BMI}}(\boldsymbol{\alpha}_{\text{parent}})$
    \ENSURE $\boldsymbol{\alpha}_{\text{opt}}$, $f^{\text{best}}_{\text{BMI}}(\boldsymbol{\alpha}_{\text{opt}})$
   
    \FOR{$i = 1$ \TO $t_{\text{s}}$}
        \STATE 
       Obtain \( \boldsymbol{L} \) as a lower triangular matrix 
       from Cholesky decomposition of the covariance matrix
\(
\boldsymbol{\Sigma} = \boldsymbol{L} \boldsymbol{L}^{\top}
\) 
        \STATE $\boldsymbol{\epsilon} = \boldsymbol{L\xi}$, where $\boldsymbol{\xi} \sim   
        \mathcal{N}(0, \mathbf{I})$
        \STATE $\boldsymbol{\alpha}_{\text{offspring}} \leftarrow \boldsymbol{\alpha}_{\text{parent}} + \sigma_{\text{loc}}\boldsymbol{\epsilon}$
        \STATE $p_\text{s} \leftarrow (1 - c_\text{p})p_\text{s} + c_\text{p} v_\text{succ}$
        \STATE $\sigma_{\text{loc}} \leftarrow \sigma_{\text{loc}} \exp{(\frac{1}{d} (p_\text{s} - \frac{p^{t}_{\text{s}}}{1 - p^{t}_{\text{s}}} (1 - p_{\text{s}})))}$
        \STATE Calculate the fitness $f_{\text{BMI}}(\boldsymbol{\alpha}_{\text{offspring}})$
        \IF{$f_{\text{BMI}}(\boldsymbol{\alpha}_{\text{offspring}}) > f_{\text{BMI}}(\boldsymbol{\alpha}_{\text{parent}})$}
            \STATE $\boldsymbol{\alpha}_{\text{parent}} \leftarrow \boldsymbol{\alpha}_{\text{offspring}}$
            \STATE Store the best solution into $\boldsymbol{\alpha}_{\text{opt}}$ and $f^{\text{best}}_{\text{BMI}}(\boldsymbol{\alpha}_{\text{opt}})$
            \IF{$p_{\text{s}} < p_{\text{t}}$}
                \STATE $\boldsymbol{p}_\text{c} \leftarrow (1 - c_{\text{cth}})\boldsymbol{p}_\text{c} + \sqrt{c_{\text{cth}}(2 - c_{\text{cth}})}\boldsymbol{\epsilon}$
                \STATE $\boldsymbol{\Sigma} \leftarrow (1 - c_{\text{cov}})\boldsymbol{\Sigma} + c_{\text{cov}}\boldsymbol{p}_{\text{c}}\boldsymbol{p}_{\text{c}}^{\top}$
            \ELSE
                \STATE $\boldsymbol{p}_\text{c} \leftarrow (1-c_{\text{cth}})\boldsymbol{p}_\text{c}$
                \STATE $\boldsymbol{\Sigma} \leftarrow (1 - c_{\text{cov}})\boldsymbol{\Sigma} + c_{\text{cov}}(\boldsymbol{p}_\text{c}\boldsymbol{p}_\text{c}^{\top} + c_{\text{cth}}(2-c_{\text{cth}})\boldsymbol{\Sigma})$
            \ENDIF
            \STATE $v_{\text{succ}} \leftarrow 1$
        \ELSE
            \STATE $v_{\text{succ}} \leftarrow 0$
        \ENDIF
    \ENDFOR
\end{algorithmic}
\end{algorithm}

Depending on \(p_{\text{s}}\), the step size is adjusted accordingly (line~6): increased to encourage broader local exploration when \(p_{\text{s}}\) is high, or reduced to promote finer granularity when \(p_{\text{s}}\) is low. Fitness evaluation follows the same rule as in~(\ref{fitness_function}), with infeasible solutions penalized by \(10^5\).

Updates to the parent vector, evolution path, and covariance matrix are conditionally applied only when an improved solution is found (line~8), which ensures search stability and directional refinement. The best local fitness is recorded in line~10 and returned to Algorithm~\ref{algorithm1} as the refined \(\boldsymbol{\alpha}^{(t)}_i\).

To further stabilize the local search, lines~11–17 of Algorithm~\ref{algorithm2} impose conditional updates to the evolution path \(\boldsymbol{p}_{\text{c}}\): if the success rate \(p_{\text{s}}\) exceeds a predefined threshold \(p_{\text{t}}\), updates are restricted to prevent overexpansion of the covariance matrix and excessive step sizes. This mechanism guards against oscillatory behavior and ensures effective fine-tuning in promising regions.

By embedding (1+1)-CMA-ES into each generation of the memetic framework, the algorithm is able to continually refine individual solutions while leveraging global information across generations, resulting in enhanced solution quality and convergence performance.

\textbf{Computational Complexity.}
For controller gain matrix 
\(\boldsymbol{F} \in \mathbb{R}^{n_u \times n_y}\), 
let \(n = n_u n_y\) denote the decision variable dimension, \(p\) the population size, and \(t_{\max}\) the number of generations.

Sampling $p
$ candidates from $\mathcal{N}(\boldsymbol{m}, \sigma^2 \boldsymbol{\Sigma})$ 
using a precomputed Cholesky factorization 
$\boldsymbol{\Sigma}^{1/2}=\boldsymbol{L}$ costs $\mathcal{O}(pn^2)$ per generation, 
and the rank-$\mu$ covariance and step–size updates cost $\mathcal{O}(n^2)$.
The Cholesky decomposition of $\boldsymbol{\Sigma}$ to refresh $\boldsymbol{L}$ costs $\mathcal{O}(n^3)$ 
when performed; if done every $k$ generations, this contributes an amortized 
$\mathcal{O}(n^3/k)$ per generation.
Thus, the global CMA-ES part contributes
\[
\mathcal{O}\!\big(t_{\max} (p n^2 + n^2 + n^3/k)\big)=\mathcal{O}\!\big(t_{\max} (p n^2 + n^3/k)\big).
\]

In the memetic step, the embedded \((1{+}1)\)-CMA-ES is invoked for at most \(r \le p\) individuals per generation, each for \(\tau\) inner iterations on average.
Each inner iteration requires one sample and rank-one update, i.e., \(\mathcal{O}(n^2)\); if a local Cholesky decomposition is done every \(k_{\mathrm{loc}}\) inner iterations, its amortized cost is \(\mathcal{O}(n^3/k_{\mathrm{loc}})\) per inner iteration.
Hence the local refinement adds
\[
\mathcal{O}\!\big(t_{\max} (r \tau n^2 + r n^3/k_{\mathrm{loc}})\big).
\]

Putting these together, Algorithm~1 has total arithmetic complexity
\[
\mathcal{O}\!\Big(t_{\max}\big[ (p + r\tau)\,n^2 \;+\; (n^3/k) \;+\; (r n^3/k_{\mathrm{loc}}) \big] \Big).
\]
In the common case where \(k,k_{\mathrm{loc}}\) are constants, this simplifies to
\(\mathcal{O}\!\big(t_{\max}\big((p + r\tau)\,n^2 + (1{+}r)\,n^3\big)\big)\).

Nevertheless, this cost is incurred entirely offline during the controller design phase and does not affect runtime performance. Once a feasible controller gain is obtained, it can be deployed without any further computation, making the offline computational cost a minor concern in practical applications.

\section{Numerical Result}\label{sec_experiment}

This section presents numerical results to illustrate our findings and summarizes the application of different methods for solving BMI-constrained optimal control problems.

The COMPleib library was used to
provide a comprehensive and diverse set of benchmark problems that are widely recognized in the control community for evaluating BMI-constrained control synthesis methods~\cite{04leibfritz11}. These benchmarks are derived from practical control problems and cover a broad range of system types, including systems with parameter uncertainty, time delays, and performance constraints.
The following abbreviations were used: academic test problems (NN), wind energy conversion (WEC), airplanes (AC), helicopters (HE), reactors (REA), decentralized interconnected systems (DIS), and terrain-following (TF).
The library includes both low-order and higher-order systems (ranging up to order 30 or more), as well as problems with challenging feasibility regions;
some problems are non-minimum phase systems (in terms of transmission zeros), such as 
AC7, HE1, HE5, NN9, and NN17.
This diversity ensures that the test cases reflect real-world control design challenges in terms of both scale and numerical complexity. Consequently, performance evaluation on the COMPleib suite provides meaningful insights into the practical robustness and scalability of proposed BMI solvers.

The evaluated methods include HIFOO~\cite{hifoo,BURKE2006339,11arzelier,09gumussoy}, LMIRank~\cite{orsilmirank}, PENBMI~\cite{05henrion}, CCDM~\cite{1206dinh}, ICAM~\cite{1212dinh}, MRVs~\cite{17chiu}, QDOM~\cite{QDOM}, standard CMA-ES~\cite{CMAES}, and the proposed memetic CMA-ES.
All simulations were performed on a workstation running Windows 11, equipped with an AMD 7955WX processor and 128 GB of DDR5 RAM. Computations were carried out in MATLAB R2024. The optimization solver used for fitness evaluation was MOSEK 10.1.15, accessed through the MOSEK Toolbox for MATLAB. The $H_\infty$-norm was computed using the MATLAB routine norm(system, r) with \(system = G_{cl}\) and \(r = \infty\).

 \subsection{Parameter Setting of Memetic CMA-ES and Comparable Methods}

The parameters in Algorithm~\ref{algorithm1}, including $c_{\sigma}$, $c_c$, $c_1$, $c_{\mu}$, and $d_{\sigma}$, follow established CMA-ES literature~\cite{BI_population_CMA-ES,hansen2016cma,Evaluating_CMA_ES}. These values are the outcome of extensive empirical tuning and theoretical justification, and have consistently demonstrated reliable performance across diverse optimization problems. Their persistent use in the literature highlights both their practical effectiveness and the robustness of CMA-ES under varying conditions.

\begin{table}[t]
\centering
\caption{Parameters of (1+1)-CMA-ES}
\begin{tabular}{|l|c|r|}
\hline
\textbf{parameter} & \textbf{symbol} &\textbf{value} \\ \hline
step size damping                  &\(d\)           & $1 + \frac{n}{2}$            \\ \hline
target success rate                &\(p^t_{\text{s}}\)       & $2/11$        \\ \hline
success rate                       &\(p_{\text{s}}\)         & $2/11$        \\ \hline
success rate averaging parameter   &\(c_{\text{p}}\)         & $1/12$     \\ \hline
cumulation time horizon parameter  &\(c_{\text{cth}}\)         & $\frac{2}{2 + n}$      \\ \hline
covariance matrix learning rate    &\(c_{\text{cov}}\)     & $\frac{2}{n^2 + 6}$  \\ \hline
threshold success rate                &\(p_{\text{t}}\)         & $0.44$  \\ \hline
step size                          & $\sigma_{\text{loc}}$         & $\frac{1}{10}\sigma^{(t)}$ \\
\hline
\end{tabular}
\label{table1}
\end{table}

Table~\ref{table1} summarizes the parameters in Algorithm~2, including $d$, $p_s^t$, $c_p$, $c_{cth}$, $c_{cov}$, and $p_t$, which were adopted from~\cite{CMAES,(1+1)_CMA_ES_cite} and validated through extensive simulations on test functions of different dimensions~\cite{(1+1)-CMAES}. These parameters were explored over wide ranges, extending beyond conventional operating limits, and ultimately selected to remain sufficiently distant from performance-critical boundaries, thereby ensuring robustness.

For $H_\infty$ optimization, both memetic CMA-ES and standard CMA-ES were executed with  $t_{\text{max}} = 10000$ to ensure fairness. To prevent excessively large controller gains, $\beta = 10^{-10}$ was used in~(\ref{fitness_function}) across all benchmarks, since larger values were observed to degrade performance. Unless otherwise specified, all other parameters followed the settings used in earlier experiments.

The QDOM algorithm was implemented with the parameter settings reported in~\cite{QDOM}. Specifically, the number of iterations for bound searching was set to $t^{\text{search-end}} = 10$, the maximum normalized search range to $r^{\text{max}} = 1$, the novelty search threshold to $r^{\text{NS-end}} = 50$, and the minimum novelty threshold to $\rho^{\text{bvr-min}} = 0.1 g^\theta$. Additional parameters included $k = 10$, $\epsilon = 0.5$, the maximum number of iterations for the immune-inspired mechanism $t^{\text{immune-max}} = 20$, $N_{\text{nom}} = 40$, and $N_{\text{max}} = 160$. The diversity evaluation coefficients were set to $\kappa_1 = 1$, $\kappa_2 = 0.5$, and $\kappa_3 = 0.1$. The population size for subspace generation was fixed at 20, with three subspaces generated per iteration. The initial value of $\lambda$ was set to 50, with $\lambda^{\text{initial}} = 20$.

For the BMI-constrained single-objective problems, the MRV algorithm followed the configuration reported in~\cite{17chiu}, with $N_{\text{nom}} = 40$, $N_{\text{max}} = 160$, and a maximum of 20 iterations.

The remaining methods, including HIFOO, PENBMI, ICAM, and CCDM, were initialized with the controller gain set to $\boldsymbol{F} = 0$, following the settings in the ICAM and CCDM studies~\cite{1206dinh,1212dinh}. The LMIRank method was implemented according to the CCDM study, where the decay rate parameter was initialized at $10^{-4}$ and increased by $0.1$ in each iteration.

\subsection{$H_\infty$ Optimization}

Table~\ref{H_infty optimization}\footnote{The numerical results for LMIRank, PENBMI, and CCDM are taken from~\cite{1206dinh}; results for ICAM are from~\cite{1212dinh}; and those for MRVs and QDOM are from~\cite{QDOM}.} presents the results of \(H_\infty\) optimization for a range of benchmark problems, comparing the performance of the proposed memetic CMA-ES algorithm—with and without the embedded refinement—to that of several existing methods, including HIFOO, PENBMI, CCDM, MRVs, and QDOM. A total of 47 BMI-constrained control problems from the COMPl\textsc{e}ib library~\cite{04leibfritz11} were evaluated. In terms of achieving the best \(H_\infty\)-norm values, the respective success rates were: HIFOO (10.64\%), PENBMI (0\%), CCDM (10.64\%), MRVs (21.28\%), QDOM (17.02\%), and the proposed memetic CMA-ES (combined results) achieved 85.11\%. 

The proposed memetic CMA-ES method achieved the best or near-best \( H_\infty \) norm in the majority of benchmark categories, including AC, HE, REA, DIS, and NN. For instance, in the AC class, our method achieved the best performance in 10 out of 13 benchmarks. In the HE and DIS problem classes, it consistently produced the best results. The only notable exception occurred in the IH benchmark, where CCDM outperformed our method. These results demonstrate that the proposed approach is broadly effective and generalizes well across diverse problem types, although further improvements may be needed to enhance performance in high-dimensional cases such as IH.

The superior performance of the proposed method can be attributed to its ability to overcome the limitations of conventional solution techniques. Traditionally, suboptimal solutions to BMI problems are obtained by alternately updating subsets of the decision variables, thereby reducing the BMI to a convex LMI problem when part of the variables is fixed. However, this approach often results in locally optimal solutions and is highly sensitive to the partitioning of variables and the choice of initial conditions. In contrast, the proposed memetic CMA-ES algorithm leverages stochastic sampling from an adaptive multivariate normal distribution, enabling exploration of a broader region of the search space.
 By reducing sensitivity to initialization and mitigating trajectory bias inherent in alternating optimization methods, the proposed approach substantially increases the likelihood of identifying high-quality solutions.

 Furthermore, the embedded local search component, implemented via the $(1+1)$-CMA-ES, performs greedy, single-solution-based refinement, which enhances exploitation once a promising region is located. This focused search strategy is particularly advantageous in BMI scenarios, where the feasible region defined by matrix inequalities is often narrow and irregular. By operating on a single candidate solution and adapting its search distribution locally, the $(1+1)$-CMA-ES is more effective at staying within these constrained regions, thereby reducing the likelihood of infeasible evaluations. In contrast, standard population-based CMA-ES may disperse samples more broadly, leading to inefficiencies in constrained environments. As shown in Table~\ref{H_infty optimization}, this additional refinement led to a performance improvement of 19.15 percentage points, calculated from the aggregated results presented in the table.

\subsection{Case Study}

We selected AC9 model (Transport Aircraft,
Boeing, cruise flight condition) for further analysis in terms of the statistical distribution of the $ \|\boldsymbol{G}_{\text{cl}} \|_\infty $ values,
computation time, and step disturbance rejection.
AC9 serves as an excellent case study because it highlights the superior performance of the proposed memetic CMA-ES, which achieved the best \(H_\infty\) result. This challenging airplane control problem is particularly illustrative as the commercial solver PENBMI failed to find any solution, thereby underscoring our method's robustness and practical applicability.

The AC9 model with $n_z=2$ and $n_w=10$ can be represented by a $2\times10$ multiple-input multiple-output
(MIMO) transfer function $\boldsymbol{G}_{\text{cl}} (s)$ as%
\[
\boldsymbol{G}_{\text{cl}} (s)=\left[
\begin{array}
[c]{ccccc}%
G_{1,1}(s) & G_{1,2}(s) & \cdots & G_{1,9}(s) &
G_{1,10}(s)\\
G_{2,1}(s) & G_{2,2}(s) & \cdots & G_{2,9}(s) &
G_{2,10}(s)
\end{array}
\right]
\]
and half of whose entries are zero transfer functions, i.e., $G_{1,i}(s)=0$
for $i=6,..10$, and $G_{2,j}(s)=0$ for $j=6,...,10$.

\begin{figure}[ptbh]
\centering
\includegraphics[width=0.8\textwidth]{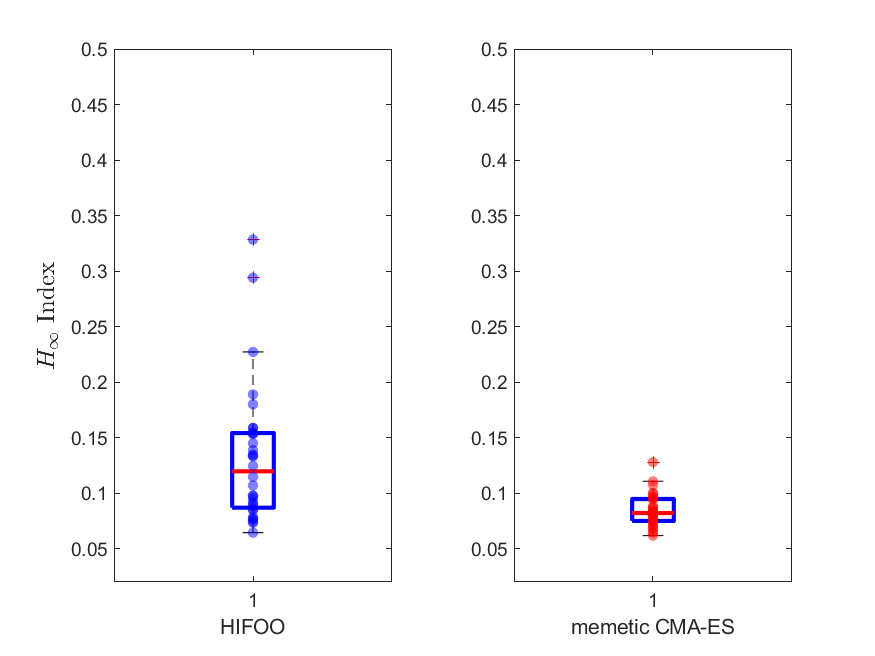}  \caption{Boxplots of
$H_{\infty}$ performance results of HIFOO and memetic CMA-ES on the AC9 model over
32 independent runs (each with 850 iterations). memetic CMA-ES achieves a lower
median with smaller variance.}%
\label{fig:H-norm}%
\end{figure}

Figure~\ref{fig:H-norm}  shows that the solutions obtained by memetic CMA-ES
exhibit a smaller standard deviation, a lower median, and fewer outliers.
These results suggest that memetic CMA-ES is more likely to converge toward
the global optimum. In the experiments, the optimal performance values of
$H_{\infty}$ performance index achieved by HIFOO and memetic CMA-ES were 0.0644
and 0.0288, respectively.

\begin{figure}[ptbh]
\centering
\includegraphics[width=0.8\textwidth]{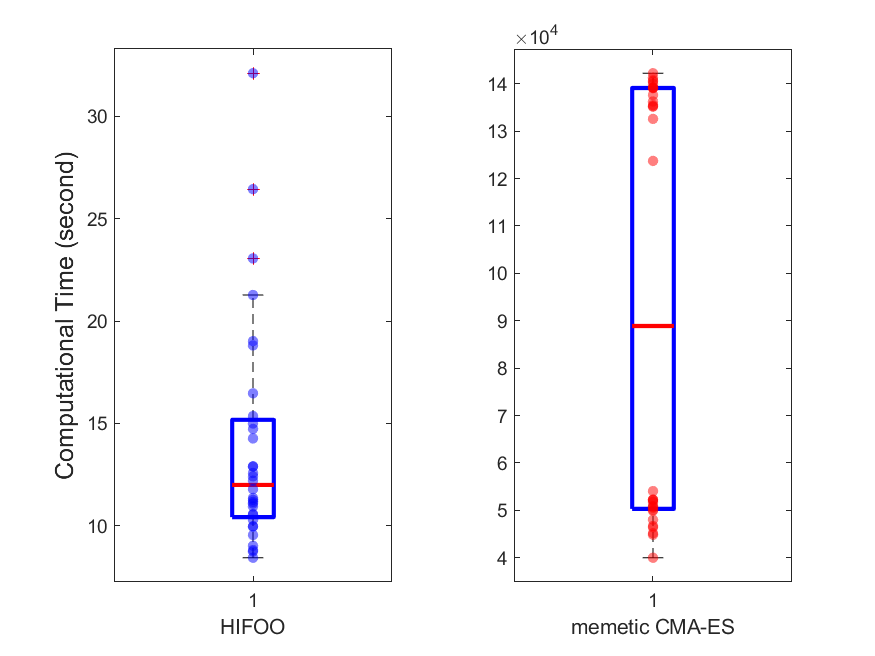}  \caption{Boxplots of
computation time for HIFOO and memetic CMA-ES on the AC9 model over 32 independent
runs. HIFOO achieves a lower median execution time with smaller variance,
while memetic CMA-ES shows longer runtime and greater variability.
It is worth noting that this computational cost is incurred entirely during the offline controller design phase and does not affect the real-time performance of the closed-loop system. Once a feasible controller gain is obtained, it can be directly implemented without further computation, making the offline cost a minor concern in practical applications.
}
\label{fig:compute-time}%
\end{figure}

 Figure~\ref{fig:compute-time} presents the boxplot of computation time. Although memetic CMA-ES
requires  more computation time than HIFOO and demonstrates
higher variability, a lower $H_{\infty}$ performance index is often desirable
in practical engineering applications, particularly for systems with limited
fault tolerance.

It is worth mentioning that this computational cost is incurred entirely during the offline controller design phase and does not affect the real-time performance of the closed-loop system. Once a feasible controller gain is obtained, it can be directly implemented without any further computation. Therefore, the offline computational cost is generally a minor concern in practical applications.

\begin{figure}[ptbh]
\hspace{-2cm}
\includegraphics[width=1.3\textwidth]{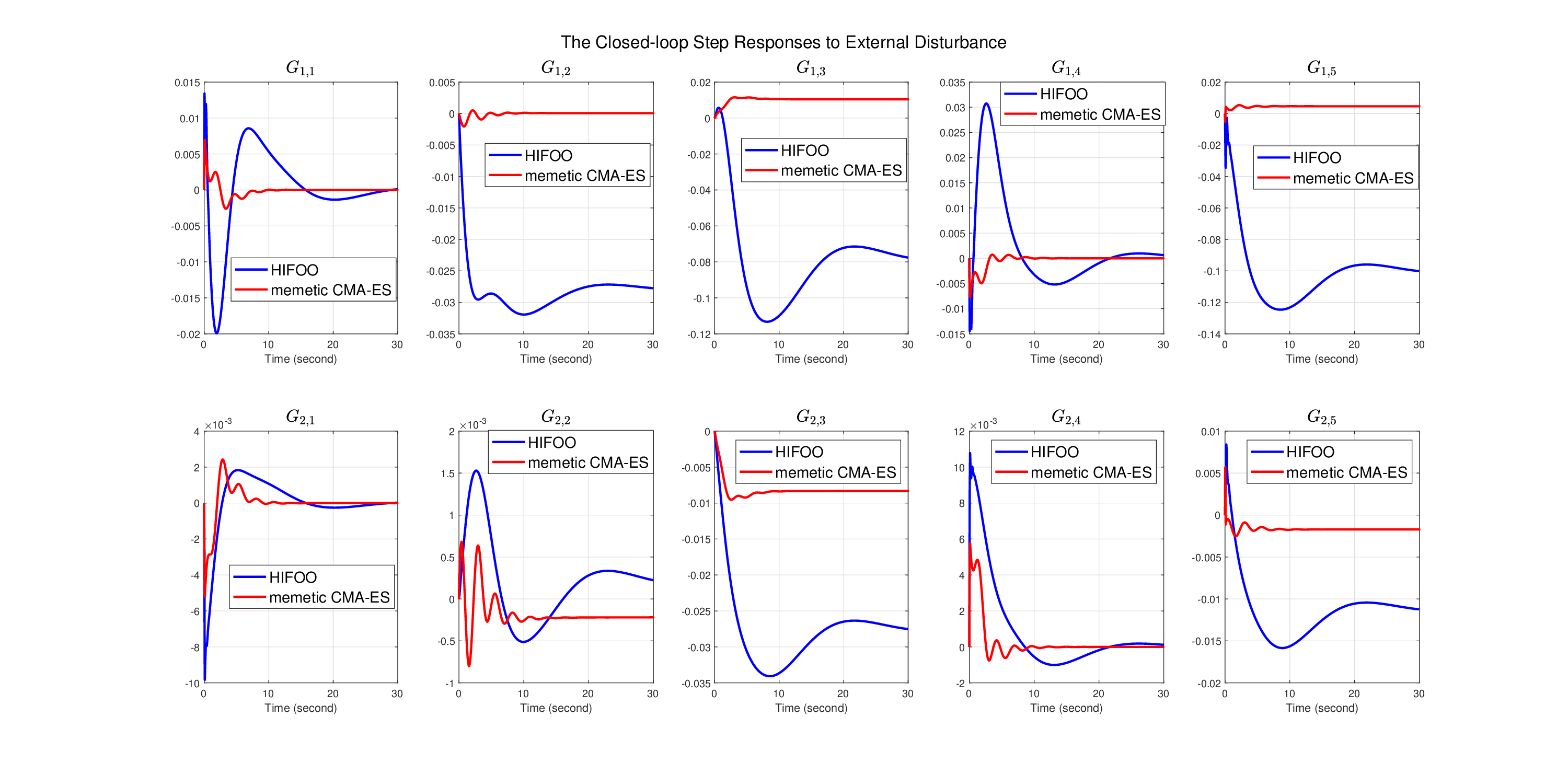}
\caption{Closed-loop step responses of the AC9 system to an exogenous step disturbance, showing that memetic CMA-ES achieves superior disturbance rejection and faster convergence compared to HIFOO.
}%
\label{fig:dis-reject}%
\end{figure}

Figure~\ref{fig:dis-reject}
shows
the system's dynamic response to a step disturbance, where only the
non-zero transfer function components are plotted. It can be
observed that memetic CMA-ES achieves superior disturbance rejection and faster
convergence. The control gains obtained from HIFOO and memetic CMA-ES,
corresponding to $H_{\infty}$ performance indices of 0.0644 and 0.0288, are
provided as follows
\[
\boldsymbol{F}_{\text{memetic CMA-ES}}=\left[
\begin{array}
[c]{ccccc}%
86.17 & -13.44 & -60.24 & 42.71 & 67.73\\
2.04 & 73.80 & 16.78 & -14.09 & -21.75\\
-21.41 & -120.66 & 23.76 & -14.44 & -20.43\\
-67.28 & 59.75 & 19.77 & -14.21 & -25.62
\end{array}
\right]
\]

\[
\boldsymbol{F}_{\text{HIFOO}}=\left[
\begin{array}
[c]{ccccc}%
3.11 & -8.74 & -8.97 & 0.04 & 8.51\\
1.12 & 28.19 & -0.01 & 0.93 & 4.46\\
-2.29 & -35.52 & 4.63 & -1.17 & -7.50\\
-1.98 & 15.41 & 4.36 & 0.20 & -5.49
\end{array}
\right]
\]

\subsection{Spectral Abscissa Optimization} 

Another commonly studied BMI-constrained control problem is spectral abscissa optimization~\cite{1206dinh}. This approach serves as a criterion for ensuring asymptotic stability of the closed-loop system. Specifically, a feedback control law guarantees asymptotic stability for the closed-loop system in~(\ref{eq:closed_loop_system}) if the optimized spectral abscissa, defined as the largest real part among all eigenvalues of the closed-loop system matrix, is strictly negative.

Table~\ref{spectral abscissa optimization} presents the results of the Spectral Abscissa Optimization across various problem instances. 
Regarding the minimization of the largest real part of the closed-loop system's eigenvalues, the observed success rates were: HIFOO (16.67\%), LMIRank (20\%), PENBMI (26.67\%), CCDM (16.67\%), ICAM (20\%), MRVs (23.33\%), QDOM (40\%), and the proposed memetic CMA-ES achieving the highest rate at 73.33\%.

\section{Conclusion}

The proposed memetic CMA-ES demonstrates a clear advantage in reliably achieving high-quality solutions across a wide range of BMI problems. Compared with existing approaches, it consistently produces superior performance and greater robustness, making it particularly effective in challenging optimization scenarios such as $H_\infty$-norm minimization and spectral abscissa reduction. While its higher computational cost is a drawback—occurring entirely during the offline design phase and not affecting real-time implementation—it may be a concern when frequent retuning or limited computational resources are involved. Nevertheless, the method is highly applicable to control problems where robustness, solution quality, and global optimality are critical. Its hybrid design, combining the global search capabilities of CMA-ES with local refinement, makes it especially suitable for nonconvex or safety-critical systems, such as aerospace applications and high-uncertainty environments, where conventional local or convex optimization methods often fail to find feasible solutions.

 As a direction for future work, the proposed approach could be extended to experimental platforms. In particular, evaluating performance under practical conditions such as sensor degradation, model mismatch, and actuation delays would provide deeper insights into the method’s applicability to real-world systems. This extension would help validate the algorithm’s robustness in the presence of hardware-related uncertainties and further enhance its relevance to safety-critical control applications.

\section{Acknowledgments}
This work was supported by the National Science and Technology Council of Taiwan under Grant 114-2221-E-305-008. 
During the preparation of this work the authors used ChatGPT  and Gemini in order to improve the clarity and fluency of the English writing. After
using this tool/service, the authors reviewed and edited the content as
needed and take full responsibility for the content of the publication.

\FloatBarrier

\begin{table}
\centering
\caption{$H_\infty$ Optimization Result}
\label{H_infty optimization}
\resizebox{1.0\textwidth}{!}{
\begin{tabular}{|r|c|c|r|r|r|r|r|r|r|r|}
\hline
\multicolumn{1}{|c|}{Problem}
&\multicolumn{2}{|c|}{$\mathbf{F} \in \mathcal{R}^{n_u \times n_y}$}
&\multicolumn{5}{|c|}{Results of existing methods, $||G_{cl}||_\infty$}
&\multicolumn{2}{|c|}{Results of CMA-ES }
\\
Name & $n_u$ & $n_y$ & HIFOO & PENBMI & CCDM & MRVs & QDOM & standard  & memetic \\
\hline
AC1 & 3 & 3 & \textbf{0.0000} & x & 0.0177 & 0.0405 & 0.0407 & 0.0182 & 0.0075\\
\hline
AC2 & 3 & 3 & \textbf{0.1115} & x & 0.1140 & 0.1262 & 0.1296 & \textbf{0.1115} & \textbf{0.1115}\\
\hline
AC3 & 2 & 4 & 4.7021 & x & \textbf{3.4859} & 3.9206 & 3.8218 & 3.6278 & 3.6278\\
\hline
AC4 & 1 & 2 & \textbf{0.9355} & x & 69.9900 & 69.99 & 69.99 & 69.99 & 69.99\\
\hline
AC6 & 2 & 4 & 4.1140 & x & 4.1954 & 4.8138 & 4.6791 & 4.1039 & \textbf{4.1015} \\
\hline
AC7 & 1 & 2 & 0.0651 & 0.3810 & 0.0548 & \textbf{0.0315} & 0.0316 & \textbf{0.0315} & \textbf{0.0315}\\
\hline
AC8 & 1 & 5 & 2.0050 & x & 3.0520 & 1.4305 & 1.4146 & \textbf{1.3849} & \textbf{1.3849}\\
\hline
AC9 & 4 & 5 & 0.0644 & x & 0.9237 & 3.2926 & 2.9749 & 0.0456 & \textbf{0.0288}\\
\hline
AC11 & 2 & 4 & 3.5603 & x & 3.0104 & 3.1158 & 2.9495 & \textbf{2.8489} & 2.8600\\
\hline
AC12 & 3 & 4 & 0.3160 & x & 2.3025 & 1.3532 & 1.2386 & \textbf{0.0438} & 0.1333\\
\hline
AC15 & 2 & 3 & 15.2074 & 427.4106 & 15.1995 & 17.1925 & 16.5535 & 16.1684 & \textbf{15.0789}\\
\hline
AC16 & 2 & 4 & 15.4969 & x & 14.9881 & 15.8600 & 15.5125 & \textbf{14.8092} & 14.8150\\
\hline
AC17 & 1 & 2 & \textbf{6.6124} & x & 6.6373 & \textbf{6.6124} & \textbf{6.6124} & \textbf{6.6124} & \textbf{6.6124}\\
\hline
HE1 & 2 & 1 & 0.1540 & 1.5258 & 0.1807 & 0.1538 & 0.1533 & \textbf{0.1526} & \textbf{0.1526}\\
\hline
HE2 & 2 & 2 & 4.4931 & x & 6.7846 & 4.3681 & 4.411 & 3.9029 & \textbf{3.8833}\\
\hline
HE3 & 4 & 6 & 0.8545 & 1.6843 & 0.9243 & 0.8570 & 0.8586 & 0.8380 & \textbf{0.8258}\\
\hline
HE4 & 4 & 6 & 23.3448 & x & \textbf{22.8713} & 46.5677 & 42.3192 & 24.0694 & 22.9887\\
\hline
HE5 & 4 & 2 & 8.8952 & x & 37.3906 & 20.8784 & 19.4099 & 8.5628 & \textbf{8.5622} \\
\hline
REA1 & 2 & 3 & 0.8975 & x & 0.8815 & 0.8836 & 0.882 & 0.8664 & \textbf{0.8657} \\
\hline
REA2 & 2 & 2 & 1.1881 & x & 1.4188 & 1.1471 & 1.1518 & 1.1480 & \textbf{1.1479} \\
\hline
REA3 & 1 & 3 & 74.2513 & 74.446 & 74.5478 & \textbf{74.2513} & \textbf{74.2513} & \textbf{74.2513} & \textbf{74.2513}\\
\hline
DIS1 & 4 & 4 & 4.1716 & x & 4.1943 & 4.3197 & 4.2662 & 4.1681 & \textbf{4.1644} \\
\hline
DIS2 & 2 & 2 & 1.0548 & 1.7423 & 1.1546 & 1.0604 & 1.0619 & 1.0481 & \textbf{1.0453} \\
\hline
DIS3 & 4 & 4 & 1.0816 & x & 1.1382 & 1.2727 & 1.2057 & \textbf{1.0809} & 1.0845 \\
\hline
DIS4 & 4 & 6 & 0.7465 & x & 0.7498 & 0.9486 & 0.8968 & 0.7406 & \textbf{0.7386} \\
\hline
TG1 & 2 & 2 & 12.8462 & x & 12.9336 & 14.2157 & 14.1969 & 12.8346 & \textbf{12.8052} \\
\hline
AGS & 2 & 2 & \textbf{8.1732} & 188.0315 & \textbf{8.1732} & 10.0239 & 10.4129 & \textbf{8.1732} & \textbf{8.1732} \\
\hline
WEC2 & 3 & 4 & 4.2726 & 32.9935 & 6.6082 & 7.8382 & 6.8686 & 4.2434 & \textbf{4.2393} \\
\hline
WEC3 & 3 & 4 & 4.4497 & 200.1467 & 6.8402 & 7.2021 & 7.3313 & 4.6277 & \textbf{4.4321} \\
\hline
BDT1 & 3 & 3 & 0.2664 & x & 0.8562 & \textbf{0.2662} & \textbf{0.2662} & \textbf{0.2662} & \textbf{0.2662} \\
\hline
MFP & 3 & 2 & 31.5899 & x & 31.6079 & 33.9193 & 33.2116 & \textbf{31.5400} & 31.5403 \\
\hline
IH & 11 & 10 & 1.9797 & x & \textbf{1.1858} & 30.1004 & 7.5915 & 2.8480 & 2.4711 \\
\hline
CSE1 & 2 & 10 & 0.0201 & x & 0.0220 & \textbf{0.0198} & 0.0199 & 0.0199 & 0.0199 \\
\hline
PSM & 2 & 3 & 0.9202 & x & 0.9227 & \textbf{0.9202} & \textbf{0.9202} & \textbf{0.9202} & \textbf{0.9202} \\
\hline
EB1 & 1 & 1 & 3.1225 & 39.9526 & 2.0276 & \textbf{1.888} & 1.8917 & 1.8917 & 1.8917 \\
\hline
EB2 & 1 & 1 & 2.0201 & 39.9547 & 0.8148 & \textbf{0.8142} & \textbf{0.8142} & \textbf{0.8142} & \textbf{0.8142} \\
\hline
EB3 & 1 & 1 & 2.0575 & 3995311.074 & 0.8153 & \textbf{0.8143} & \textbf{0.8143} & \textbf{0.8143} & \textbf{0.8143} \\
\hline
NN1 & 1 & 2 & 13.9782 & 14.6882 & 18.4813 & 15.5294 & 14.1901 & 13.8291 & \textbf{13.7442} \\
\hline
NN2 & 1 & 1 & 2.2216 & x & \textbf{2.2216} & 2.2038 & \textbf{2.2216} & \textbf{2.2216} & \textbf{2.2216} \\
\hline
NN4 & 2 & 3 & 1.3627 & x & 1.3802 & 1.4327 & 1.3985 & \textbf{1.3586} & 1.3594 \\
\hline
NN8 & 2 & 2 & 2.8871 & 78281181.15 & 2.9345 & 2.9193 & 2.9254 & 2.8853 & \textbf{2.8849} \\
\hline
NN9 & 3 & 2 & 28.9083 & x & 32.1222 & 30.7173 & 30.7665 & \textbf{27.6047} & 27.6122 \\
\hline
NN11 & 3 & 5 & 0.1037 & x & 0.1566 & 0.1075 & 0.1023 & \textbf{0.0919} & 0.0920 \\
\hline
NN15 & 2 & 2 & 0.1039 & x & 0.1194 & \textbf{0.098} & \textbf{0.098} & \textbf{0.098} & \textbf{0.098} \\
\hline
NN16 & 4 & 4 & 0.9557 & x & 0.9656 & 2.3044 & 2.1322 & 0.9544 & \textbf{0.9478} \\
\hline
NN17 & 2 & 1 & 11.2182 & x & 11.2381 & 11.2042 & 11.2031 & \textbf{11.2011} & \textbf{11.2011} \\
\hline
\multicolumn{9}{l}{The letter ``x'' means that no solution was found, and the best value is highlighted in bold.} \\
\end{tabular}
}
\end{table}

\begin{table}[t]
\caption{SPECTRAL ABSCISSA OPTIMIZATION}
\label{spectral abscissa optimization}
\resizebox{\textwidth}{!}{
\begin{tabular}{|r|c|c|r|r|r|r|r|r|r|r|}
\hline
Name & $n_u$ & $n_y$ & HIFOO & LMIRank & PENBMI & CCDM & ICAM & MRVs & QDOM & Memetic CMA-ES  \\
\hline
AC1 & 3 & 3 & -0.2061 & -8.4766 & -7.0758 & -0.8535 & -0.7814 & -18.0761 & -52.2891 & \textbf{-191.1893} \\
\hline
AC4 & 1 & 2 & \textbf{-0.05} & \textbf{-0.05} & \textbf{-0.05} & \textbf{-0.05} & \textbf{-0.05} & \textbf{-0.05} & \textbf{-0.05} & \textbf{-0.05} \\
\hline
AC5 & 2 & 2 & -0.7746 & -1.8001 & -2.0438 & -0.7389 & -0.7389 & -2.4051 & \textbf{-2.4159} & -1.6223 \\
\hline
AC7 & 1 & 2 & -0.0322 & -0.0204 & 0.0896 & -0.0673 & -0.0502 & -0.0747 & -0.0872 & \textbf{-0.0902} \\
\hline
AC8 & 1 & 5 & -0.1968 & \textbf{-0.4447} & \textbf{-0.4447} & -0.0755 & -0.0640 & \textbf{-0.4447} & \textbf{-0.4447} & \textbf{-0.4447} \\
\hline
AC9 & 4 & 5 & -0.3389 & -0.5230 & -0.4450 & -0.3256 & -0.3926 & -2.0823 & \textbf{-2.3951} & -1.3890 \\
\hline
AC11 & 2 & 4 & -0.0003 & -5.0577 & x & -3.0244 & -3.1573 & -16.9018 & -138.8328 & \textbf{-140.8926} \\
\hline
AC12 & 3 & 4 & -10.8645 & -9.9658 & -1.8757 & -0.3414 & -0.2948 & -18.3236 & -52.5179 & \textbf{-296.1378} \\
\hline
HE1 & 2 & 1 & -0.2457 & -0.2071 & \textbf{-0.2468} & -0.2202 & -0.2134 & -0.2446 & -0.2462 & -0.2370 \\
\hline
HE3 & 4 & 6 & -0.4621 & -2.3009 & -0.4063 & -0.8702 & -0.8380 & -1.7847 & \textbf{-3.2928} & -1.0198 \\
\hline
HE4 & 4 & 6 & -0.7446 & -1.9221 & -0.0909 & -0.8647 & -0.8375 & -3.0567 & \textbf{-3.0968} & -1.2392 \\
\hline
HE5 & 4 & 2 & -0.1823 & x & -0.2932 & -0.0587 & -0.0609 & \textbf{-1.1953} & -1.0151 & -0.6973 \\
\hline
HE6 & 4 & 6 & \textbf{-0.0050} & \textbf{-0.0050} & \textbf{-0.0050} & \textbf{-0.0050} & \textbf{-0.0050} & \textbf{-0.005} & \textbf{-0.005} & \textbf{-0.005} \\
\hline
REA1 & 2 & 3 & -16.3918 & -5.9736 & -1.7984 & -3.8599 & -2.8932 & -19.3041 & -199.4919 & \textbf{-304.5592} \\
\hline
REA2 & 2 & 2 & -7.0152 & -10.0292 & -3.5928 & -2.1778 & -1.9514 & -19.4238 & -29.6645 & \textbf{-68.1247} \\
\hline
REA3 & 1 & 3 & \textbf{-0.0207} & \textbf{-0.0207} & \textbf{-0.0207} & \textbf{-0.0207} & \textbf{-0.0207} & \textbf{-0.0207} & \textbf{-0.0207} & \textbf{-0.0207} \\
\hline
DIS2 & 2 & 2 & -6.8510 & -10.1207 & -8.3289 & -8.4540 & -8.3419 & -19.4340 & -68.9765 & \textbf{-119.9200} \\
\hline
DIS4 & 4 & 6 & -36.7203 & -0.5420 & -92.2842 & -8.0989 & -5.4467 & -16.0222 & -185.1219 & \textbf{-355.7095} \\
\hline
WEC1 & 3 & 4 & -8.9927 & -8.7350 & -0.9657 & -0.8779 & -0.8568 & -11.9629 & -21.6390 & \textbf{-16.0217} \\
\hline
IH & 11 & 10 & \textbf{-0.5000} & \textbf{-0.5000} & \textbf{-0.5000} & \textbf{-0.5000} & \textbf{-0.5000} & -0.1576 & -0.3516 & \textbf{-0.5000} \\
\hline
CSE1 & 2 & 10 & -0.4509 & -0.4844 & -0.4490 & -0.2360 & -0.2949 & -0.3489 & -0.4372 & \textbf{-0.5056} \\
\hline
TF1 & 2 & 4 & x & x & -0.0618 & -0.1544 & -0.0704 & -0.2688 & -0.2573 & \textbf{-0.2862} \\
\hline
TF2 & 2 & 3 & x & x & \textbf{-1.0e-5} & \textbf{-1.0e-5} & \textbf{-1.0e-5} & \textbf{-1.0e-5} & \textbf{-1.0e-5} & \textbf{-1.0e-5} \\
\hline
TF3 & 2 & 3 & x & x & \textbf{-0.0032} & -0.0031 & \textbf{-0.0032} & \textbf{-0.0032} & \textbf{-0.0032} & \textbf{-0.0032} \\
\hline
NN1 & 1 & 2 & -3.0458 & -4.4021 & -4.3358 & -0.8746 & 0.1769 & -5.8900 & -5.8928 & \textbf{-5.9102} \\
\hline
NN5 & 1 & 2 & -0.0942 & -0.0057 & -0.0942 & -0.0913 & -0.0490 & -0.0940 & \textbf{-0.0942} & \textbf{-0.0942} \\
\hline
NN9 & 3 & 2 & -2.0789 & -0.7048 & x & -0.0279 & 0.0991 & -17.8516 & \textbf{-25.8536} & -22.9967 \\
\hline
NN13 & 2 & 2 & -3.2513 & -4.5310 & -9.0741 & -3.4318 & -0.2783 & -13.6061 & -13.6067 & \textbf{-15.2267} \\
\hline
NN15 & 2 & 2 & -6.9983 & \textbf{-11.0743} & -0.0278 & -0.8353 & -1.0409 & -10.9821 & -10.9741 & -8.2159 \\
\hline
NN17 & 2 & 1 & \textbf{-0.6110} & -0.5130 & x & -0.6008 & -0.5991 & -0.6107 & -0.6107 & \textbf{-0.6110} \\
\hline
\multicolumn{11}{l}{The letter ``x'' means that no solution was found, and the best value is highlighted in bold.} \\
\end{tabular}
}
\end{table}

\FloatBarrier

\appendix
\renewcommand{\thesection}{Appendix}
\renewcommand{\theequation}{\arabic{equation}} 
\setcounter{equation}{0} %
\renewcommand{\thefigure}{\arabic{figure}}
\section{}
\label{appendix:BMI_constraints}
Let \(n=n_u  n_y\).
The following parameters  were set:
\begin{equation*}
    p = 4 + \left\lfloor 3 \ln n \right\rfloor \quad
\end{equation*}
\begin{equation*}
    \label{weight}
    \theta_i = \ln \frac{  5 + \left\lfloor 3 \ln n \right\rfloor  }{2} - \ln i \quad \text{for } i = 1, \dots, \mu
\end{equation*}
The expected Euclidean norm of a random vector following a standard multivariate normal distribution \( \mathcal{N}(\boldsymbol{0}, \mathbf{I}) \), where the mean vector is zero and the covariance matrix is the identity, can be approximated by:
\begin{equation*}
    \mathbb{E}\|\mathcal{N}(\boldsymbol{0}, \mathbf{I})\| \approx \sqrt{n} \left( 1 - \frac{1}{4n} + \frac{1}{21n^2} \right)
\end{equation*}
This approximation estimates the average length of a vector sampled from an \( n \)-dimensional standard normal distribution.

\begin{align*}
    \mu_{\text{eff}} = \frac{(\sum^{\mu}_{i = 1}|\theta_i|^2)^2}{\sum^{\mu}_{i = 1}\theta^2_i} = \frac{1}{\sum^{\mu}_{i = 1}\theta^2_i}
\end{align*}
\begin{align*}
    (\boldsymbol{\Sigma}^{(t)})^{-\frac{1}{2}} \overset{\mathrm{def}}{=} \boldsymbol{B}^{(t)} \boldsymbol{D}^{(t)^{-1}} \boldsymbol{B}^{(t)^\top}
\end{align*}
In the diagonal matrix $\boldsymbol{D}^{(t)}$, the diagonal elements are square roots of the corresponding positive eigenvalues and $\boldsymbol{B}^{(t)}$ is an orthonormal basis of eigenvectors. The parameter settings were primarily derived from extensive simulations conducted on basic test functions across various dimensions~\cite{hansen2023cmaevolutionstrategytutorial}:
\begin{align*}
    c_\sigma &= \frac{\mu_\text{eff} + 2}{n + \mu_\text{eff} + 5}\\
    c_{\text{c}} &= \frac{4 + \mu_\text{eff} / n}{n + 4 + 2 \mu_\text{eff} / n} \\
    h_\sigma &= \left\{
    \begin{array}{ll}
        1 & \mbox{if } \frac{\|p_\sigma\|}{\sqrt{1 - (1 - c_\sigma)^{2(g+1)}}} < \left(1.4 + \frac{2}{n+1}\right) \mathbb{E} \| \mathcal{N}(\mathbf{0}, \mathbf{I}) \| \\
        0 & \mbox{otherwise}
    \end{array}
    \right. \\
    d_\sigma &= 1 + 2 \max \left(0, \sqrt{\frac{\mu_\text{eff} - 1}{n + 1} - 1}\right) + c_\sigma \\
    c_1 &= \frac{2}{(n + 1.3)^2 + \mu_\text{eff}} \\
    c_\mu &= \min \left(1 - c_1, \frac{1 / 2 + 2\mu_\text{eff} + 2 / \mu_\text{eff} - 4}{(n + 2)^2 + \mu_\text{eff}}\right)
\end{align*}

\begin{spacing}{1}
\bibliographystyle{elsarticle-num}
\bibliography{elsbib_applied_soft_computing}
\end{spacing}
\end{document}